
\input harvmac
\magnification=1200
\Title{\vbox{\baselineskip12pt\hbox{IC/93/237}}}
{\vbox{\centerline{Extended Nonabelian Symmetries}
\vskip2pt\centerline{for Free Fermionic Model}}}

\centerline{Raiko P. Zaikov{\footnote{$^\dagger $}{Permanent
address: Institute for
Nuclear Recearch and Nuclear Energy, Boul. Tzarigradsko Chaussee
72, 1784 Sofia, Bulgaria, e-mail: zaikov@bgearn.bitnet}}
\footnote{$^\star $}{Financial suport
from Bulgarian Fundamental Recearch Fondation under contract
Ph-11, 91-94}}
\bigskip\centerline{International Centre for Teoretical Physics}
\centerline{Strada Costeiera 11, P. O. Box 11, Trieste, Italy}
\centerline{}

\vskip .3in

\noindent {\bf Abstract} The higher spin symmetry for both Dirac
and Majorana massless free fermionic field models are
considered. An infinite Lie algebra which is a linear realization
of the higher spin extension of the cross products of the
Virasoro and affine Kac-Moody algebras is obtained. The
corresponding current algebra is closed which is not the case
of annalogous current algebra in the WZNW model. The gauging
procedure for the higher spin symmetry is given also.

\Date{6/93}

\newsec{Introduction}

\noindent  In the paper \ref\rW{E.  Witten, Comm. in Math. Phys.,
{\bf 92}, 455 (1984).}, it was shown that the WZNW model admits
the same symmetries $O(N)\otimes O(N)$, Virasoro and Kac-Moody on
the classical level as (in the critical point) on the quantum
level. A similar result was obtained in the papers \ref\rPW{A.
Polyakov and P.  B. Wiegmann, Phys. Lett., {\bf 131B}, 121
(1983).} and
\ref\rPWi{A.  Polyakov and P. B. Wiegmann, Phys.  Lett., {\bf
141B}, 223 (1984).} where the gauged fermionic model and the
gauged WZNW model were considered. In the last few years, higher
spin extensions of the Virasoro algebra were under intensive
considerations (for a complete list of references see \ref\rBS{P.
Boukwnegt and K.  Schoutens, {\it "W-Symmetry in Conformal Field
Theory"}, CERN preprint CERN-TH/6583/92, to be published in Phys.
Reports.}).  Recently, the higher spin extended symmetries for
the nonabelian free massive Majorana fermionic model have been
investigated in the papers
\ref\rSS{G. Sotkov and  M. Stanishkov, {\it "Off-critical
$W_\infty $ and Virasoro algebras as dynamical symmetries of the
integrable models"}, Preprint IFT - P 001/93, Sao Paulo,
hep-th/9301066} and
\ref\rAASS{E. Abdalla, M. C. B. Abdalla, G. M. Sotkov and M.
Stanishkov, {\it "Of critical current algebra"},
IFUSP-preprint-1027, Sao-Paolo 1993; hep-th/9302002}. The
corresponding current algebra contains the $W_\infty $ current
algebra as a subalgebra.  A similar problem was considered in
Ref. \ref\rRZ{R. P. Zaikov {\it Noabelian higher spin extended
symmetries for the classical WZNW model} Preprint INRNE TH/1/93;
hep-th/9303087} where it was shown that the classical WZNW model
admits nonlinearly realized $w_\infty $ symmetry, both linearly
and nonlinearly realized $W_\infty$ symmetry  as well as linearly
realized extended affine Kac-Moody symmetry.  However, the
current algebras which correspond to the latter symmetries
(except the $w_\infty $) are not closed.  The later makes it
impossible to gauge (see Refs.
\ref\rH{C. M. Hull, Phys. Lett. {\bf B240}, 110 (1990); {\bf
B259}, 68 (1991); Nucl. Phys., {\bf B353}, 707 (1991); {\bf
B364}, 621 (1991).} and \ref\rBPSS{E. Bergshoeff, C. N. Pope, L.
J. Romans, E.  Sezgin, X. Shen and K. S. Stelle, Phys. Lett.,
{\bf B243}, 350 (1990).}) these symmetries in the case of the
WZNW model.  Consequently, there arrises a difference between the
nonabelian free fermionic model and the WZNW model on the
extended symmetry level.

The main goal of the present article is to make a further
clarification of the above mentioned difference between
two-dimensional free fermionic model and the corresponding WZNW
model. Here we investigate the higher spin extended symmetries for
the nonabelian $SO(N)$, \ $SL(N)$ and $SU(N)$ free fermionic
models in the Lagrangian approach. It is shown that all these
symmetries admit  gauging, because on both classical and quantum
levels there is $W_\infty $ symmetry as well as extended
Kac-Moody symmetry whose currents form an invariant space. The
latter and the results from \rRZ \ show that the nonabelian free
fermionic model and the WZNW model are not equivalent on the
higher spin symmetry level (see \rSS , \rAASS , \rRZ \ and
\ref\rRZa{R. P. Zaikov, {\it "Higher spin symmetries in WZNW
model"}, in preparation}). The symmetries considered here are
off-shell, however, the corresponding Noether currents are
nonhermitean in the basis which is choosen for convenience.  This
nonhermitean form of the currents simplify the derivation of the
Lie algebra in both the classical and quantum cases as well as
the obtaining of the transformation laws for the gauge fields.
Due to the fact that not all of the currents in the Majorana
case are independent, there arises a symmetry of Stuckelberg
type \rBPSS . Moreover, in this nonhermitean basis the central terms are
nondiagonal with respect to the conformal spin which means that
the transition amplitudes between the states with different spins
are nonzero, i.e. we are dealing with nonphysical basis. We note,
that in this case the Lie algebra and the corresponding current
algebra have more simple forms than the corresponding algebra in
the physical basis (with diagonal central terms) \ref\rP{C. N.
Pope, L. J. Romands and X. Schen, Phys. Lett., {\bf B236}, 173 (1990)
and {\bf B242}, 401 (1990).}, \ \rAASS .  The transition to
the physical basis  can be made
by simple redefinition of the currents which induces trivial
deformation of the Lie algebra as well as of the current
algebra.

\newsec{Classical free fermionic algebra}

\noindent In order to show the difference between the WZNW model
and the nonabelian free fermionic model we
consider higer spin extension of the Virasoro and of the affine
Kac-Moody algebras in the case of free complex fermionic
realization. Higher spin extension of the $O(N)$ Kac-Moody algebra
for the massive free Majorana spinor feild was obtained in Ref.
\rSS \ and \rAASS . In the present article we will show that this
extension has a more simple form for the complex spinor fields than
for the Majorana ones.

In the case  of $SU(N) \ {\rm or} \ SL(N,C) $ free
fermionic model, we have a complex spinor field for which the free
field action reads:
\eqn\a{S=\int dzd\bar z\bigl(\bar \psi _+\partial _{\bar z}\psi _++
\bar \psi_-\partial _z\psi _-\bigr)}
It is easy to verify that this action is off-shell invariant with
respect to the following extended conformal and extended
global (semilocal) nonabelian transformations:
\eqn\b{\eqalign{\delta ^m\{k\}\psi (z) & =
k_m(z)\partial ^{m+1}\psi (z), \cr
\delta ^m\{k\}\bar \psi (z) & =
(-)^{m+2}\partial ^{m+1}\bigl(k_m(z)\bar \psi (z)\bigr),
\qquad (m=0, 1, \dots ),\cr }}
\eqn\c{\eqalign{\widehat \delta ^m[\hat \alpha ]\psi (z) & =
\alpha ^a_mt_a\partial ^m\psi (z), \cr
\widehat \delta ^m[\hat \alpha ]\bar \psi (z) & =
(-)^{m+1}\partial ^m\bigl(\alpha ^a_m(z)\bar \psi (z)t_a\bigr),
\qquad (m=0, 1, \dots ) \cr }}
where $k(z)$ and $\alpha (z)$ are arbitrary holomorphic
functions for the $\psi_+$--component and antiholomorphic for the
$\psi_-$--component\foot{In the Minkowsky space-time
$k$ and $ \alpha $ are arbitrary function of the corresponding single
light-cone variable only} and $t_a$ are generators of $SU(N)$ or
$SL(N,C)$ in fundamental representation. In what follows we will
consider only the $\psi_+$--component, keeping in mind that the
$\psi_-$--component has quite similar behavior.

 In the case when $k$ and $\alpha $ are arbitrary
holomorphic functions the transformations $\delta ^0\psi $ and
$\widehat \delta _a^0\psi $ form the classical (without
central term) Virasoro algebra and the classical affine Kac-Moody
algebra. For arbitrary $m$ (arbitrary spin) we derive
the following Lie algebra:

\eqn\efca{\bigl[\delta ^m\{k\},\delta ^n\{h\}\bigr]\psi (z)=
\sum _{p=0}^{max(m+1,n+1}\delta ^{m+n-p+1}\{[k_m,h_n]^p_-\}\psi
(z),}
\eqn\efcb{
\bigl[\delta ^m\{k\},\widehat \delta ^n_a\{\alpha \}\bigr]\psi (z)=
\sum _{p=0}^{max(m+1,n}\widehat \delta _a^{m+n-p+1}\{[k_m,\beta
^a_{n-1}]^p_-\}\psi (z),}
\eqn\efcc{\eqalign{
\bigl[\widehat \delta _a^m\{\alpha ^a\},\widehat \delta _b^n\{\beta ^b\}
\bigr]\psi (z) & ={1\over 2}
\sum _{p=0}^{max(m,n}\biggl(f_{ab}^c\widehat \delta _c^{m+n-p+1}
\{[\alpha ^a_m,\beta ^b_{n-1}]^p_+\}\psi(z) \cr
& +d_{ab}^c\widehat \delta _c^{m+n-p+1}
\{[\alpha ^a_m,\beta ^b_{n-1}]^p_-\}\psi(z) \cr
& +2\delta ^{m+n-p+1}
\{[\alpha ^a_m,\beta ^a_{n-1}]^p_+\}\psi(z)\biggr), \cr }}
where
\eqn\eww{[k_m,h_n]^p_{\pm }=\pmatrix{ n+1 \cr p \cr }h_n\partial
^k_m \pm
\pmatrix{ m+1 \cr p \cr }\delta ^{m+n-p+1}k_m\partial ^ph_n,}

$f_{abc}=tr([t_a,t_b]t_c), d_{abc}=tr(\{t_a,t_b\}t_c)$,
the matrix generators $t_a$ are normalized by
$tr(t_at_b)=2\delta _{ab}$ and  is taken into account that the
binomial coefficients $\pmatrix{m \cr p \cr }=0$ for $p>m$. We
note that for the derivation of \efcc \ the identity
\eqn\efcd{t_at_b={1\over 2}(f_{abc}+d_{abc})t_c+2\delta _{ab}I,}
is used also. We remind, that this identity is satisfyied for the
$SU(N)$ and $SL(N,R)$ generators in fundamental representation
but it is not saisfyied for the $SO(N)$ generators in the adjoint
representation. This property of the $SU(N)$ generators  makes
its higher spin extension more simple than the higher spin
extension of the $SO(N)$ algebra (see \rSS \ and \rAASS ).

The Lie algebra \efca \ -- \efcc \ coincides with the obtained in
\rRZ \ Lie algebra for $SU(N)$ WZNW model.

{}From \efca \ -- \efcc \ it follows that the extension of the
$SU(N)$ ($SL(N,C)$) Kac-Moody algebra is not closed.
 To obtain a closed algebra we have to start with
both Virasoro and Kac-Moody algebras, i.e. to consider the higher
spin extension of the $U(N)$ ($GL(N,C)$) algebra.

The conserved Neother currents corresponding to \b \ and \c \
are given by:
\eqn\efba{V^m(z)=\bar \psi \partial ^{m+1}\psi (z), }
\eqn\efbb{J^m_a(z)=\bar \psi t_a\partial ^m\psi (z). }
These currens are written in a nonsymmetric form and therefore
they are nonhermitean which leads to the appearance of
nondiagonal central terms in the OPE.  This form, however is more
convenient for applications. A transition to more symmetric form
of \b \ and \c \ can be carried out by the following simple
redefinition
\eqn\efbc{\delta ^m\psi \rightarrow \widetilde \delta ^m\psi =
\sum _{l=0}^mC^m_l\partial ^lk_m\partial ^{m-l}\psi .}
This redefinition leads to a redefinition of the coresponding
current
\eqn\efbd{U^m\rightarrow \widetilde U^m=
\sum _{l=0}^m(-)^lC^m_l\partial ^lU^{m-l}\psi .}
It is clear that the redefinition of the transformations leads to
 a deformation of the
algebra \efca \ -- \efcc \ also. Under suitable choice of the
coefficients $C$ in \efbc \ the algebra \efca \
coincides with the $W_\infty $ algebra (see \rP ).

To demonstrate the difference with the WZNW model we will
show that the classical currents \efba \ and \efbb \ form an
invariant space with respect to the transformations \b \ and
\c . Namely
\eqn\efda{\eqalign{\delta ^l\{k\}V^m(z) &
=-\sum _{p=0}^{l+1}\sum _{q=0}^{l-p+1}(-)^{p+q}\pmatrix{ l+1 \cr
p \cr }\pmatrix{ l-p+1 \cr q \cr }\partial ^pk_l\partial ^q
V^{l+m-p-q+1} \cr & + i\sum _{p=0}^{m+1}\pmatrix{ m+1 \cr p \cr
}\partial ^pk_lV^{l+m-p+1}, \cr
\delta ^l\{k\}J_a^m(z) &
=-\sum _{p=0}^{l+1}\sum _{q=0}^{l-p+1}(-)^{p+q}\pmatrix{ l+1 \cr
p \cr }\pmatrix{ l-p+1 \cr q \cr }\partial ^pk_l\partial ^q
J_a^{l+m-p-q+1} \cr & + i\sum _{p=0}^m\pmatrix{ m \cr p \cr
}\partial ^pk_lJ_a^{l+m-p+1}, \cr  }}
\eqn\efdc{\eqalign{\widehat \delta _a^l\{k\}V^m(z) &
=-\sum _{p=0}^l\sum _{q=0}^{l-p}(-)^{p+q}\pmatrix{ l \cr p \cr
}\pmatrix{ l-p \cr q \cr }\partial ^pk_l\partial ^q
J_a^{l+m-p-q+1} \cr & + i\sum _{p=0}^m\pmatrix{ m+1 \cr p \cr
}\partial ^pk_lJ_a^{l+m-p+1}, \cr
\widehat \delta _a^l\{\alpha \}J_b^m(z) &
={1\over 2}(f_{abc}-d_{abc})
\sum _{p=0}^l\sum _{q=0}^{l-p}(-)^{p+q}\pmatrix{ l \cr p \cr
}\pmatrix{ l-p \cr q \cr }\partial ^p\alpha _l\partial ^q
J_c^{l+m-p-q} \cr & +
{1\over 2}(f_{abc}+d_{abc})
\sum _p^m\pmatrix{ m \cr p \cr }\partial ^p\alpha _l^b
J_c^{l+m-p} \cr
& -\delta _{ab}\sum _{p=0}^l\sum _{q=0}^{l-p}(-)^{p+q}\pmatrix{ l \cr p \cr
}\pmatrix{ l-p \cr q \cr }\partial ^p\alpha _l\partial ^q
V^{l+m-p-q-1} \cr & +
\delta _{ab}\sum _p^m\pmatrix{ m \cr p \cr }\partial ^p\alpha _l
V_c^{l+m-p-1} \cr }}

In spite of the fact that the field transformation laws form the
Lie algebra \efca \ -- \efcc \ which coincides whith those in the
WZNW model, the current transformation laws are different. In the
WZNW model the
classical higher spin bilinear currents, except of $w_\infty $
nonlinear currents,  do not form a closed current algebra \rRZ \
which is not the case for the laws \efda \ and \efdc .

We note, that from \efda \ it follows
\eqn\efdd{\eqalign{\delta ^0V^m & =
(m+2)\partial k_0{\cal V}^m+k_0\partial V^m+
\sum _{p=2}^{m+1}\pmatrix{ m+1 \cr p \cr }\partial ^pk_0
V^{m-p+1}, \cr
\delta ^0J_a^m & =
m\partial k_0J_a^m+k_0\partial J_a^m+
\sum _{p=2}^m\pmatrix{m \cr p \cr }\partial ^pk_0
J_a^{m-p+1}, \cr }}
which shows that if $m>0$ the higher spin
energy-momentum tensors and currents transform with respect to
quasiprimary transformation law.

\newsec{Free fermionic operator algebra}

\noindent  We find the quantum
conserved currents from the classical ones \efba \ and
\efbb \ applying a
suitable normal ordering prescription:
\eqn\or{\eqalign{{\cal V}^m(z) & =:\bar \psi \partial ^{m+1}\psi :,
\cr
{\cal J}_a^m(z) & =:\bar \psi t_a\partial ^m\psi :.
\cr }}
Following Schoutens et all. \ref\rBBS{F. A. Bais, P. Bouwknegt,
M. Surridge and K. Schoutens, Nucl. Phys., {\bf B304}, 348
(1988).} we define
\eqn\os{:A(z)B(z):={1\over 2\pi i}\oint _{\Gamma }{dx\over x-z}
A(x)B(z),}
where $\Gamma $ is a small countur arround the point $z$.

If we substitute $\bar \psi =\psi $ in \or \ we will obtain the
Majorana fermion currents. In this case there arizes isotopic
symmetric traseless tensor conserved current \rAASS
\eqn\eor{J^m_{ab}(z)=:\psi t_{ab}\partial ^{m+1}\psi :,}
where
\eqn\eora{t_{ab}={1\over 2}\{t_a,t_b\}-2\delta _{ab}I.}

The Dirac and Majorana spinor cases we consider separately.

\subsec{Dirac spinor field}

\noindent Applying
the ordering prescription we obtain the singular terms of the
operator product expansion of two currents \or :
\eqn\eob{\eqalign{{\cal V}^k(z){\cal V}^l(w) & \sim
\sum _{p=0}^{k+1}\sum _{q=0}^pP^k_{pq}
(z-w)^{p-k-2}\partial ^{p-q}{\cal V}^{l+q}(w) \cr
& +\sum _{p=0}^{l+1}
{(l+1)!\over p!}(z-w)^{p-k-l-3}{\cal V}^{k+p+1}(w)
+2NC^D_{k,l}, \cr }}
where
\eqn\eoba{P^m_{pq}=(-)^{m+q-1}{(m+1)!\over (p-q)!q!}, \qquad \
C^D_{mn}=2(-)^m(m+1)!(n+1)!.}
{}From \eob \ and \eoba \ it follows that nondiagonal central
charges appear. In the same way we derive
\eqn\eod{\eqalign{{\cal V}^k(z){\cal J}^l_a(w) & \sim
\sum _{p=0}^{k+1}\sum _{q=0}^pP^k_{pq}
(z-w)^{p-k-2}\partial ^{p-q}{\cal J}_a^{l+q}(w) \cr
& +\sum _{p=0}^{l}{l!\over
p!}(z-w)^{p-l-1}{\cal J}_a^{k+p+1}(w) \cr }}
and
\eqn\eoe{\eqalign{{\cal J}_a^k(z){\cal J}_b^l(w) & \sim
{f_{abc}+d_{abc}\over 2}\sum _{p=0}^k\sum _{q=0}^pP^{k-1}_{pq}
(z-w)^{p-k-1}\partial ^{p-q}{\cal
J}_c^{l+q}(w) \cr
& +{f_{abc}-d_{abc}\over 2}\sum _{p=}^{l}{l!\over p!}
(z-w)^{p-l-1}{\cal J}_c^{k+p}(w) \cr
& +\delta _{ab}\biggl(\sum _{p=0}^k\sum _{q=0}^pP^{k-1}_{pq}
(z-w)^{p-k-1}
\partial ^{p-q}{\cal
V}^{l+q-1}(w) \cr
& +\sum _{p=0}^{l}{l!\over p!}(z-w)^{p-k-l-1}{\cal
V}^{k+p-1}(w)\biggr) \cr
& -2C^D_{k-1,l-1}\delta _{ab}. \cr }}
 Here the identity
 \eqn\eor{\partial ^l\bar \psi A\partial ^m\psi =
 \sum _{p=0}^l(-)^{l-p}\pmatrix{l \cr p \cr }\partial ^p
 \biggl(\bar \psi \partial ^{m+l-p}\psi \biggr), }
where $A$ is a constant matrix, is used essentially. This identity
is a consequence of the Leibniz formula. It is easy to see that
in the case $k=l=0$ we obtain from
\eob --- \eoe \ the Virasoro --- Kac-Moody operator algebra. For
$k=0$ \ and \ $l>0$ the qusiprimary transformation laws for the
quantum $\cal V$ and $\cal J$ are obtained.

The last terms of \eob \ and \eoe \ show that the central terms
are nondiagonal in the basis into consideration. Moreover, if the
lowest spin is one then the central terms form a degenerate
matrix.

Applying the operator product technique we derive quantum
transformation laws:
\eqn\eof{\eqalign{\delta ^m\psi (z) & ={1\over 2\pi i}\oint {dx\over
x-z}k_m(x)\psi (z){\cal V}^m(x)=k_m(z)\partial ^{m+1}\psi (z),
\cr
\delta ^m_a\bar \psi (z) & ={1\over 2\pi i}\oint {dx\over
x-z}k_m(x){\cal V}^m(x)=
(-)^{m+1}\partial ^{m+1}\bigl(k_m(z)\bar \psi (z)\bigr), \cr
\widetilde \delta ^m\psi (z) & ={1\over 2\pi i}\oint {dx\over
x-z}\alpha _m(x)\psi (z){\cal J}_a^m(x)
=\alpha _m(z)\partial ^mt_a\psi (z),
\cr
\widetilde \delta_a ^m\bar \psi (z) & ={1\over 2\pi i}\oint {dx\over
x-z}\alpha _m(x){\cal J}_a^m(x)\bar \psi (z)=
(-)^m\partial ^{m+1}\bigl(\alpha _m(z)\bar \psi (z)\bigr)
\cr }}
The form of these laws coincide  with the form of the
corresponding classical laws
\b \ and \c .  In the same way we can obtain the quantum
transformation laws for the currents \efba \ and \efbb \ which
coincide with the corresponding classical transformation laws
\efda -\efdd .

\subsec{Majorana spinor field}

\noindent  In the case of Majorana spinor field applying the
ordering prescription \or \ we obtain the following singular
terms in the product of two Majorana spinor currents:
\eqn\ot{\eqalign{\widehat V^m(z)\widehat V^n(w)
& \sim
\sum _{p=0}^{m+1}\sum _{q=0}^p\widetilde P^m_{pq}(z-w)^{p-m-2} \partial
^q\widehat V^{n+p-q}(w) \cr
& -\sum _{p=0}^{n+1}{(n+1)!\over p!}
(z-x)^{p-m-n-3}\widehat V^{m+p}(w) \cr
& +\sum _{p=0}^{m+n+2}Q^{mn}_p(z-w)^{p-m-n-3}\widehat V^{p-1}(w)
\cr
& -(z-w)^{-1}\sum _{q=0}^{m+1}R^m_q
\partial ^q\widehat V^{m+n-q+1}(w)
+NC_{m,n} \cr }}
The comparison of this formula with \eob \ allows to conclude
that in the Majorana case addional terms appears. These
terms are a consequence of the fact that for the Majorana fields
there are extra pairings compared to the Dirak case. Further we
obtain
\eqn\ou{\eqalign{\widehat V^m(z)\widehat J_a^n(w)
& \sim
\sum _{p=0}^{m+1}\sum _{q=0}^p\widetilde P^m_{pq}
(z-w)^{p-m-2} \partial ^q\widehat J_a^{n+p-q}(w)
-\sum _{p=0}^{n}{n!\over p!}(z-x)^{p-n-1}
\widehat J_a^{m+p+1}(w) \cr
& +\sum _{p=0}^{m+n+1}\widetilde P^{m,n-1}_p(z-w)^{p-m-n-2}\widehat J_a^p(w)
-(z-w)^{-1}\sum _{q=0}^{m+1}R^m_q
\partial ^q\widehat J_a^{m+n-q}(w) \cr }}
\eqn\ov{\eqalign{\widehat J_a^m(z)\widehat J_b^n(w)
& \sim {1\over 2}f_{ab}^c\biggl(
\sum _{p=0}^m\sum _{q=0}^p\widetilde P^{m-1}_{pq}
(z-w)^{p-m-1} \partial ^q\widehat J_c^{n-p-q}(w) \cr
& +\sum _{p=0}^{n}{n!\over p!}(z-x)^{p-n-1}\widehat
J_c^{m+p}(w)
-\sum _{p=0}^{m+n}Q^{m-1,n-1}_p(z-w)^{p-m-n-1}\widehat J_c^p(w) \cr
& -(z-w)^{-1}\sum _{q=0}^{m}R^{m-1}-q
\partial ^q\widehat J_c^{m+n-q}(w)\biggr) \cr
& +\sum _{p=0}^m\sum _{q=0}^p\widetilde P^{m-1}_{pq}
(z-w)^{p-m-1} \partial ^q\widehat J_{ab}^{n+p-q}(w)
-\sum _{p=0}^{m+n}{n!\over p!}(z-x)^{p-n-1}\widehat
J_{ab}^{m+p}(w) \cr
& +\sum _{p=0}^{m+n}Q^{m-1,n-1}_p
(z-w)^{p-m-n-1}\widehat J_{ab}^p(w)
-(z-w)_{-1}\sum _{q=0}^{m}R^{m-1}_q
\partial ^q\widehat J_{ab}^{m+n-q}(w) \cr
& -{2\delta _{ab}\over N}\biggl(\sum _{p=0}^m
\sum _{q=0}^p\widetilde P^{m-1}_p
(z-w)^{p-m-1} \partial ^q\widehat V^{n+p-q-1}(w) \cr
& -\sum _{p=0}^{n}{n!\over p!}(z-x)^{p-n-1}\widehat
V^{m+p-1}(w)
+\sum _{p=0}^{m+n}Q^{m-1,n-1}_q
(z-w)^{p-m-n-1}\widehat V^{p-1}(w) \cr
& -(z-w)^{-1}\sum _{q=0}^{m}R^{m-1}_q
\partial ^q\widehat V^{m+n-q-1}(w)\biggr)
+C^M_{m-1,n-1}\delta _{ab} ,\cr }}
where $t_{ab}$ are given by \eora \ and
\eqn\ota{\eqalign{\widetilde P^m_{pq} &
=(-)^{m+p-q+1}{(m+1)!\over
(p-q)!q!}, \cr
Q^{mn}_p & =(-)^{m+1}{(m+n+2)!\over p!}, \cr
R^m_p & =(-)^{m-p+1}\pmatrix{m+1 \cr p \cr} \cr
C^M_{m,n} &
=2N(-)^{m+n}{(m+1)!(n+1)!-(m+n+2)!\over (z-w)^{m+n+4}}. \cr }}
We note that in the case of
$SO(3)$--algebra $t_{ab}$ have the following simple form
\eqn\aua{(t_{ab})_j^k=\delta _{aj}\delta _b^k-\delta _a^k\delta _{bj}-
{2\over 3}\delta _{ab}I.}
Next we derive
\eqn\aub{\eqalign{\widehat V^m(z)\widehat J_{ab}^n(w) & \sim
\sum _{p=0}^{m+1}\sum _{q=0}^p\widetilde P^m_{pq}
(z-w)^{p-m-2} \partial ^q\widehat J_{ab}^{n+p-q+1}(w) \cr
& -\sum _{p=)}^{n+1}{(n+1)!\over p!}
(z-x)^{p-m-n-3}\widehat J_{ab}^{p-1}(w) \cr
& +(-)^{m+1}\sum _{p=0}^{m+n+2}Q^{m,n}_p
(z-w)^{p-m-n-3}\widehat J_{ab}^{p-1}(w) \cr
& +(z-w)^{-1}\sum _{q=0}^{m+1}R^m_q
\partial ^q\widehat J_{ab}^{m+n-q+1}(w), \cr }}
\eqn\anc{\eqalign{\widehat J_a^m(z)\widehat J_{bc}^n(w) & \sim
D_{a,bc}^d\biggl(\sum _{p=0}^m\sum _{q=0}^p\widetilde P^{m-1}_p
(z-w)^{p-m-1} \partial ^q\widehat J_d^{n+p-q+1}(w) \cr
& +\sum _{p=0}^{m}{n!\over p!}(z-x)^{p-n-2}\widehat
J_d^{m+p}(w)
-\sum _{p=0}^{m+n}R^{m-1,n}_p(z-w)^{p-m-n-2}\widehat
J_d^p(w) \cr
& -(z-w)^{-1}\sum _{q=0}^{m}Q^{m-1}_q
\partial ^q\widehat J_d^{m+n-q+1}(w)\biggr) \cr
& +F_{a,bc}^{de}\biggl(\sum _{p=0}^m\sum
_{q=0}^p\widetilde P^{m-1}_{pq}(z-w)^{p-m-1} \partial
^q\widehat J_{de}^{n+p-q}(w) \cr
& -\sum _{p=0}^{n+1}{(n+2)!\over
p!}(z-x)^{p-n-2}\widehat J_{de}^{m+p}(w) \cr
& +(-)^{m}\sum
_{p=0}^{m+n+1}R^{m-1,n}_p(z-w)^{p-m-n-2}\widehat J_{de}^p(w) \cr
& -(z-w)^{-1}\sum _{q=0}^{m}Q^{m-1}_q
\partial ^q\widehat J_{de}^{m+n-q}(w)\biggr) \cr }}
Here the
coefficients \ $D_{a,bc}^d$ \ and \ $F_{a,bc}^{de}$ \ are
determined from the equation
\eqn\aud{t_at_{cd}=D_{a,bc}^dt_d+F_{a,bc}^{de}t_{de},}
where we take into account that \ $t_a$ \ and \ $t_{ab}$ \ form a
complete basis in the space of real traceless \ $N\times N$ \
matrices. We
note, that \ $D$ \ are symmetric, i.e. \ $D_{a,bc}^d=D_{bc,a}^d$
\ while \ $F$ \ are antisymmetric, i.e. \
$F_{a,bc}^{de}=-F_{bc,a}^{de}$.

Further we obtain also
\eqn\aue{\eqalign{\widehat J_{ab}^m(z)\widehat J_{cd}^n(w) & \sim
{\cal F}_{ab,cd}^e\biggl(\sum _{p=0}^{m+1}\sum _{q=0}^p\widetilde P^m_p
(z-w)^{p-m-2} \partial ^q\widehat J_e^{n+p-q+1}(w) \cr
& +\sum _{p=0}^{n+1}{(n+1)!\over p!}(z-x)^{p-n-2}\widehat
J_e^{m+p+1}(w) \cr
& -\sum _{p=0}^{m+n+2}Q^{m,n}_p(z-w)^{p-m-n-3}\widehat J_e^p(w) \cr
& -(z-w)^{-1}\sum _{q=0}^{m+1}R^m_q
\partial ^q\widehat J_e^{m+n-q+2}(w)\biggr) \cr
& +{\cal D}_{ab,cd}^{eg}
\biggl(\sum _{p=0}^{m+2}\sum _{q=0}^p\widetilde P^m_{pq}
(z-w)^{p-m-2} \partial ^q\widehat J_{eg}^{n+p-q+1}(w) \cr
& -\sum _{p=0}^{n+1}{(n+1)!\over p!}(z-x)^{p-n-2}\widehat
J_{eg}^{m+p+1}(w) \cr
& +\sum _{p=0}^{m+n+2}Q^{m,n}_p
(z-w)^{p-m-n-3}\widehat J_{eg}^p(w) \cr
& -(z-w)^{-1}\sum _{q=0}^{m+1}R^m_q
\partial ^q\widehat J_{eg}^{m+n-q+2}(w)\biggr) \cr
& +\delta _{ab}^{cd}\biggl(\sum _{p=0}^{m+1}
\sum _{q=0}^p\widetilde P^m_{pq}
(z-w)^{p-m-2} \partial ^q\widehat V^{n+p-q-2}(w) \cr
& -\sum _{p=0}^{m+n+2}{(n+1)!\over p!}(z-x)^{p-n-2}\widehat
V^{m+p-2}(w) \cr
& +(-)^{m}\sum _{p=0}^{m+n+2}Q^{m,n}_p
(z-w)^{p-m-n-3}\widehat V^{p-1}(w) \cr
& -(z-w)^{-1}\sum _{q=0}^{m+1}R^m_q
\partial ^q\widehat V^{m+n-q-2}(w)\biggr)
8+C^M_{m,n}\delta _{ab}^{cd} \cr }}
Here the coefficients \ $ {\cal F} $ \ and \ ${\cal D}$ \ are
determined from the equation
\eqn\aug{t_{ab}t_{cd}={\cal F}_{ab,cd}^et_e+
{\cal D}_{ab,cd}^{ef}t_{ef}+\delta _{ab}^{cd}I}
where \ ${\cal F}_{ab,cd}^e=-{\cal F}_{cd,ab}^e,
{\cal D}_{ab,cd}^{ef}={\cal D}_{cd,ab}^{ef} \ {\rm and} \
\delta _{ab}^{cd}=tr(t_{ab}t^{cd})$ .

Applying the operator product technique we derive the following
quantum transformation laws:
\eqn\auh{\eqalign{\delta ^m\psi (z) &
={1\over 2\pi i}\oint dxk_m(x)\widehat V^m(x)\psi (z) \cr
& =-k_m(z)\partial ^{m+1}\psi (z)+(-)^{m+1}\partial ^{m+1}\bigl(
k_m(z)\psi (z)\bigr), \cr
\tilde \delta ^m\psi (z) &
={1\over 2\pi i}\oint dx\alpha_m^a(x)\widehat J_a^m(x)\psi (z) \cr
& =-\alpha ^a_m(z)t_a\partial ^m\psi (z)+
(-)^m\partial ^m\bigl(\alpha ^a_m(z)\psi (z)\bigr)t_a, \cr
\widehat \delta ^m\psi (z) &
={1\over 2\pi i}\oint dx\alpha_m^{ab}(x)\widehat J_{ab}^m(x)\psi (z) \cr
& =-\alpha ^{ab}_m(z)t_{ab}\partial ^{m+1}\psi (z)+
(-)^{m+1}\partial ^{m+1}\bigl(\alpha ^{ab}_m(z)
\psi (z)\bigr)t_{ab}, \cr  }}
It is easy to check that these transformations are off-shell
symmetry of the Majorana spinor action.

\newsec{Gauging the extended affine symmetry}

\noindent We consider also the problem of gauging the
classical symmetry
corresponding to the transformations \b \ and \c \  and the
corresponding quantum symmetry \eof . First we consider the case of
$SU(N)$ fermionic model. In this case the Neother coupling with
currents \efba \ and \efbb \ is given by
\eqn\ega{L_{int}={\cal A}^a_m{\cal J}^m_a+{\cal H}_m{\cal V}^m}
For simplicity we restrict our considerations only to the case
of chiral gauge in which ${\cal A}_m$ are gauge fields only with
$m+1$  antiholomorphic indeces ($\bar z$) and ${\cal H}_m$
are gauge fields with $m+2$ antiholomorphic indeces.

{}From the invariance of the total action with respect to local
gauge transformations
corresponding to \b \ and \c \
we obtain the following transformation laws for the gauge
potentials ${\cal A}_m \ {\rm and} \ {\cal H}_m$:
\eqn\egb{\eqalign{\delta {\cal A}^a_m & =
\sum _{l\ge 0}\biggl(\sum _{p=0}^{l+1}(-)^{l+1}
\pmatrix{ l+1 \cr p \cr }
k_l\partial _z^p{\cal A}^a_{m-l+p-1}
 -\sum _{p=0}^l\pmatrix{l \cr p \cr }
\partial _z^pk_{m-l+p-1}{\cal A}^a_l\biggr) \cr
\delta {\cal H}_m & =-\partial _{\bar z}k_m+
\sum _{l\ge 0}\sum _{p=0}^l\pmatrix{ l+1 \cr p \cr }\biggl(
(-)^{l+1}k_l\partial _z^p{\cal H}_{m-l+p-1}
 -\partial _z^pk_{m-l+p-1}{\cal H}_l\biggr) \cr }}
\eqn\egc{\eqalign{\widehat \delta {\cal A}^a_m & =-\partial _{\bar
z}\alpha _m^a
+{1\over 2}\sum _{l\ge 0}\sum _{p=0}^l\pmatrix{ l+1 \cr p \cr }\biggl(
(-)^{l+1}(f_{abc}-d_{abc})
\alpha _l^b\partial _z^p{\cal A}^c_{m-l+p-1} \cr
& +(f_{abc}+d_{abc})\partial _z^p\alpha ^b_{m-l+p-1}{\cal
A}^a_l\biggr) \cr
\widehat \delta {\cal H}_m & =
\sum _{l\ge 0}\biggl(\sum _{p=0}^l(-)^l\pmatrix{ l \cr p \cr }
\alpha ^a_l\partial _z^p{\cal A}^a_{m-l+p-1}
 -\sum _{p=0}^{l+1}\pmatrix{l+1 \cr p \cr }
\partial _z^p\alpha ^a_{m-l+p-1}{\cal A}^a_l\biggr) \cr }}

As we shall see below the gauge
transformation laws in the case of $SU(N)$ fermionic model also have
a more simple form than in the Majorana spinor model. We note
also, that all the currents \efba \ and \efbb \ are
independent and consequently in the case of $SU(N)$ fermionic
model the Stukelberg like symmetry \rBPSS \ is abcent.

In order to gauge
the $SO(N)$ nonabelian Majorana spinor
theory we include a Noether coupling to the free field
action \a :
\eqn\eh{L_{int}= A^a_mJ^m_a+B_mV^m+A^{ab}_mJ^m_{ab},}
where  $V, J_a \ {\rm and} \ J_{ab}$ \ are the spinor field
currents  and $A_a, A_{ab} \ {\rm and} \ B$ \ are gauge fields
(in chiral gauge).
The transformation laws for the gauge filds we determine from the
invariance of the total action with respect to local gauge
transformations corresponding to \auh .  We note, that we obtain
the current transformation laws from the singular
terms of the OPE \ot - \ov , \aub , \anc
\ and \aue \ applying the formulas:
\eqn\eha{\eqalign{\delta ^mO(z) & =
{1\over 2\pi i}\oint dxk_m(x)V^m(x)O(z), \cr
\widehat \delta ^mO(z) & =
{1\over 2\pi i}\oint dx\alpha^a_m(x)J_a^m(x)O(z), \cr
\widetilde \delta ^mO(z) & =
{1\over 2\pi i}\oint dx\alpha ^{ab}_m(x)J_{ab}^m(x)O(z), \cr }}
As an example we give the explicite form of the most complicated law
\eqn\ehb{\eqalign{\widehat \delta J^m_b(z) & \approx
{f_{ab}^c\over 2}\biggl(\sum _{p=0}^m\sum _{q=0}^p
(-)^{m+p-q}{m!\over (p-q)!q!(m-p)!}
\partial ^{m-q}\alpha ^a_m(z)\partial ^qJ^{n+p-q}_c(z) \cr
& +\sum _{p=m}^{m+n}{n!\over (p-m)!(n-p)!}\partial ^{n-p}
\alpha ^a_m(z)J^{m+p}_c(z) \cr
& -\sum _{p=o}^{m+n}{(m+n)!\over p!(m+n-p)!}\partial ^{m+n-p}
\alpha ^a_m(z)J^p_c(z)
-\alpha ^a_m(z)\sum _{q=0}^m\pmatrix{m \cr q \cr }
\partial ^qJ^{m+n-q}_c(z)\biggr) \cr
& +\sum _{p=0}^m\sum _{q=0}^p(-)^{m+p-q}{m!\over (p-q)!q!(m-p)!}
\partial ^{m-q}\alpha ^a_m(z)\partial ^qJ^{n+p-q}_{ab}(z) \cr
& +\sum _{p=m}^{m+n}{n!\over (p-m)!(n-p)!}\partial ^{n-p}
\alpha ^a_m(z)J^{m+p}_{ab}(z) \cr
& -\sum _{p=o}^{m+n}{(m+n)!\over p!(m+n-p)!}\partial ^{m+n-p}
\alpha ^a_m(z)J^p_{ab}(z)
-\alpha ^a_m(z)\sum _{q=0}^m\pmatrix{m \cr q \cr }
\partial ^qJ^{m+n-q}_{ab}(z) \cr
& -{2\over N}\biggl(\sum _{p=0}^m\sum _{q=0}^p
(-)^{m+p-q}{m!\over (p-q)!q!(m-p)!}
\partial ^{m-q}\alpha ^b_m(z)\partial ^qV^{n+p-q-1}(z) \cr
& +\sum _{p=m}^{m+n}{n!\over (p-m)!(n-p)!}\partial ^{n-p}
\alpha ^a_m(z)V^{m+p-1}(z) \cr
& -\sum _{p=o}^{m+n}{(m+n)!\over p!(m+n-p)!}\partial ^{m+n-p}
\alpha ^a_m(z)V^{p-1}_c(z)
-\alpha ^a_m(z)\sum _{q=0}^m\pmatrix{m \cr q \cr }
\partial ^qV^{m+n-q-1}(z)\biggr) \cr }}

As follows from \eha \ the currents \or \ form a
closed space with respect to the gauge transformations \auh . The
latter allows the theory to be gauged. The same follows for the
classical currens \efba \ \efbb \ (see the formulas \efda \ and
\efdc . In such a way we obtain the following transformation laws
for the gauge fields:
\eqn\ehz{\eqalign{\delta h_l(z) & =-\partial _{\bar z}k_l(z,\bar z)+
\sum _{\ge 0}\sum _{p=0}^{m+1}\biggl({(m+1)!\over (m-p+1)!}
\sum _{q=0}^p{(-)^{m+p}\over (p-q)!q!}\partial ^q\bigl(h_{l-p+q}
\partial ^{m-p+1}k_m\bigr)\cr
& +{1\over p!}h_m\partial _{m-p+1}k_{l-p}-
(-)^m\pmatrix{m+1 \cr q \cr }\partial ^q\bigl(k_mh_{l-m+q}\bigr)\biggr)
\cr
& +\sum _{m=0}^{l+1}\sum _{n\ge l-m-1}{(m+n+2)!\over (l+1)(m+n-l+1)!}
h_n\partial ^{m+n-l+1}k_m, \cr
\delta A^a_l(z) & =
\sum _{\ge 0}\sum _{p=0}^{m+1}\biggl({(m+1)!\over (m-p+1)!}
\sum _{q=0}^p{(-)^{m+p}\over (p-q)!q!}\partial ^q\bigl(A^a_{l-p+q}
\partial ^{m-p+1}k_m\bigr)\cr
& +{1\over (m+1)p!}A^a_m\partial _{m-p}k_{l-p}-
(-)^m\pmatrix{m+1 \cr q \cr }\partial ^q\bigl(k_mA^a_{l-m+q}\bigr)\biggr)
\cr
& +\sum _{m=0}^{l+1}\sum _{n\ge l-m-1}{(m+n+2)!\over (l+1)(m+n-l+1)!}
A^a_n\partial ^{m+n-l}k_m, \cr
\delta C^{ab}_l & =\delta (A^a\longrightarrow C^{ab}) \cr }}

The remaining transformation laws have a similar form. We note
that the nonzero nonhomogenious terms are only the following:
\eqn\ehu{\widehat \delta A^a_l\approx -\partial _{\bar z}\alpha ^a_l,
\quad \quad \widetilde \delta A^{ab}_l\approx
-\partial _{\bar z}\alpha ^{ab}_l.}

In the Majorana case as a concequence of \efba , \efbb \ and \eor
\ only the even spin energy-momentum tensors,
the even spin isotopic tensor currents and
the odd spin currents are independent. This statement is a
consequence of the Leibniz formula \eor \ from which we get
\eqn\aza{U^{2k} =
-{1\over 2}\sum_{p=1}^{2k+1}(-)^{p}\pmatrix{2k \cr p
\cr }\partial ^pU^{2k-p},}
for symmetric matrix $A$ and
\eqn\azb{J_a^{2k+1}=-{1\over 2}\sum _{p=1}^{2k+1}(-)^p
\pmatrix{2k+1 \cr p \cr }\partial ^p J_a^{2k-p+1}}
for antisymmetric $A$. Here $U^m=V^{m}$ or $J^m_{ab}$. These
fomulas allow us to compute the coefficients in
\eqn\azc{U^{2k+1}=\sum _{m=0}^kC_m^k\partial ^{2m+1}U^{2(k-m)}.}
Then the formula \azc
\ shows  that the interaction action \eh \ admits a Stuckelberg
type symmetry \rBPSS . This symmetry express in:
\eqn\ehv{\eqalign{\Delta A^a_{2m+1} &
=\chi ^a_m, \cr
\Delta A^a_{2m} & =
\sum _{\ge }C_p^{m+p}\partial ^{2p+1}\chi ^a_{m+p}, \cr
\Delta B_{2m+1} & =\eta _m, \cr
 \Delta B_{2m} & =
\sum _{\ge )}C_p^{m+p}\partial ^{2p+1}\eta _{m+p}, \cr
\Delta A^{ab}_{2m+1} & =\eta ^{ab}_m, \cr
\Delta ^{ab}_{2m} & =
\sum _{\ge }C_p^{m+p}\partial ^{2p+1}\eta ^{ab}_{m+p}, \cr }}
where $\eta $ are arbitrary functions. This invariance allows us
to choose the following (additional to the chiral gauge) gauge
fixing:
\eqn\ehve{B_{2m+1}=0, \qquad \ A^a_{2m+1}=0, \qquad \
A^{ab}_{2m+1}=0.}
In this gauge the even spin gauge fields $B_{2m+1}$,
$A^{ab}_{2m+1}$, and the odd spin isotopic vector potential $A^a$
are
canceled in the Lagrangian \eh .

The one-loop two-particle vertex function in the Dirac spinor
case is given by:
\eqn\epa{\Gamma_{(2)}\approx \sum _{m,n\ge 0}\int dzd\bar z
\biggl({\cal A}^a_m(z,\bar z){\partial _z^{m+n+1}
\over \partial _{\bar z}}{\cal A}^a_n(z,\bar z)+
{\cal H}_m(z,\bar z){\partial _z^{m+n+3}
\over \partial _{\bar z}}{\cal H}_n(z,\bar z)\biggr).}
The nondiagonality of the two-particle vertex function is a
consequence of the basis which we use.
Taking into account \egb \ and \egc \ we obtain
\eqn\epb{\eqalign{\delta \Gamma _2({\cal A}, {\cal H }) &\approx
\sum _{m,n\ge 0}\int dzd\bar z
\biggl({\cal H}_m(z,\bar z)\partial _z^{m+n+3}k_n(z,\bar z)+
\partial _z^{m+n+3}k_m(z,\bar z){\cal H}_n(z,\bar z)\biggr) \cr
\widetilde \delta \Gamma _2({\cal A}, {\cal H }) &\approx
\sum _{m,n\ge 0}\int dzd\bar z
\biggl({\cal A}^a_m(z,\bar z)\partial _z^{m+n+1}\alpha ^a_n(z,\bar z)+
\partial _z^{m+n+1}\alpha ^a_m(z,\bar z){\cal A}^a_n(z,\bar z)\biggr) \cr }}

Consequently, the vertex function $\Gamma _2({\cal A}_m,{\cal
A}_n)$ is invariant with respect to $SL(m+n+1)$ , while the
vertex function $\Gamma _2({\cal H}_m,{\cal H}_n)$ is invariant
with respect to $SL(m+n+3)$. We obtain  similar results
in the Majorana spinor case.

We remind that in the case of WZNW model bilinear conserved
currents exist, however, they do not form an invariant space and
consequently this higher spin simmetry cannot be gauged except the
ordinary Virasoro-Kac-Moody symmetry which corresponds to $m=n=0$.
The latter shows that on higher spin level there is no
equivalence between the nonabelian free fermionic model
and the WZNW model.


\noindent {\bf Acnowledgements.}

\noindent The author would like to thank to Prof. Abdus Salam,
the International Atomic Energy Agency and UNESCO for
hospitality in ICTP in Trieste where the present article was
completed and to Dr. L. Nikolova for a critical reading of the
manuscript.

\listrefs
\bye